\documentclass [12pt,a4paper]{article}
\usepackage{amssymb, theorem}
\newtheorem{lemma}{Lemma}
\newtheorem{prop}{Proposition}
\newtheorem{thm}{Theorem}
\newtheorem{coro}{Corollary}
\theorembodyfont{\rm}
\newtheorem{ex}{Example}
\def\Ae{\mathcal A}
\def\A0{\mathcal A_0}
\def\Pe{\mathcal P}
\def\qed{{\hfill $\square$}}
\def\abs{< \!\! <}
\def\fel{{\textstyle {1 \over 2}}}
\newcommand{\<}{\langle}
\renewcommand{\>}{\rangle}
\def\im{\rm{i}}
\def\iM{\mathcal M}
\def\iA{\mathcal A}
\def\iB{\mathcal B}
\def\iH{\mathcal H}
\def\iK{\mathcal K}
\def\iN{\mathcal N}
\def\iP{\mathcal P}
\def\iS{\mathcal S}
\def\iX{\mathcal X}

\def\vfi{\varphi}
\def\aa{\alpha}
\def\bbbn{{\mathbb N}}
\def\bbbr{{\mathbb R}}
\def\bbbc{{\mathbb C}}

\def\supp{\mbox{supp}\,}
\def\Tr{\mathrm Tr}

\def\sig{\sigma}
\def\ssum{\textstyle{\sum}}
\def\dint{\int_\oplus}
\begin{document}
\ \vskip 1cm
\centerline{\LARGE Sufficiency in quantum statistical inference}
\bigskip
\centerline{\today}
\bigskip
\bigskip
\centerline{\large Anna Jen\v cov\'a\footnote{Supported by the EU
Research Training Network Quantum Probability with Applications to Physics,
Information Theory and Biology. E-mail: jenca@mat.savba.sk.}}
\centerline{Mathematical Institute of the}
\centerline{Slovak Academy of Sciences}
\centerline{Stefanikova 49, Bratislava, Slovakia}
\bigskip
\centerline{\large D\'enes Petz\footnote{Supported by the Hungarian grant 
OTKA T032662. E-mail: petz@renyi.hu.}}
\centerline{Alfre\'ed R\'enyi Institute of Mathematics}
\centerline{Hungarian Academy of Sciences}
\centerline{POB 127, H-1364 Budapest, Hungary}
\bigskip

\medskip\bigskip

\begin{quote}
This paper attempts to develop a theory of sufficiency in the setting
of non-commutative algebras parallel to the ideas in classical 
mathematical statistics. Sufficiency of a coarse-graining means that 
all information is extracted about the mutual relation of a given
family of states. In the paper sufficient coarse-grainings are 
characterized in several equivalent ways and the non-commutative 
analogue of the factorization theorem is obtained. Among the 
applications the equality case for the strong subadditivity of the
von Neumann entropy, the Imoto-Koashi theorem and exponential
families are treated. The setting of the paper allows the underlying
Hilbert space to be infinite dimensional.
\end{quote}

\begin{quote}
MSC:  46L53, 81R15, 62B05. 
\end{quote}

\begin{quote}
{\it Key words: Quantum statistics, coarse-graining, factorization theorem, 
exponential family, strong subadditivity of entropy, sufficient subalgebra} 

\end{quote}
\bigskip\bigskip
\noindent

\section{Introduction and preliminaries}
A quantum mechanical system is described by a C*-algebra, the dynamical
variables (or observables) correspond to the self-adjoint elements and
the physical state of the system are modelled by the normalized positive
functionals of the algebra, see \cite{BEH, BR}. The evolution of the
system $\iM$ can be described in the {\bf Heisenberg picture} in which
an observable $A \in \iM$ moves into $\alpha(A)$, where $\alpha$ is a
linear transformation. $\alpha$ is an automorphism in case of the time
evolution of a closed system but it could be the irreversible evolution
of an open system. The {\bf Schr\"odinger picture} is dual, it gives
the transformation of the states, the state $\vfi \in \iM^*$ moves
into $\vfi \circ \alpha$. The algebra of a quantum system is typically
non-commutative but the mathematical formalism brings commutative algebras
as well. A simple {\bf measurement} is usually modelled by a family of pairwise
orthogonal projections, or more generally, by a partition of unity,
$(E_i)_{i=1}^n$. Since all $E_i$ are supposed to be positive and
$\sum_i E_i=I$, $\beta: \bbbc^n \to \iM$, $(z_1,z_2,\dots, z_n)\mapsto
\sum_i z_i E_i$ gives a positive unital mapping from the commutative
C*-algebra $\bbbc^n$ to the non-commutative algebra $\iM$.
Every positive unital mapping occurs in this way. The essential concept
in quantum information theory is the state transformation which is affine
and the dual of a positive unital mapping. All these and several other
situations justify to study of positive unital mappings between C*-algebras
from a quantum statistical viewpoint.

If the algebra $\iM$ is ``small'' and $\iN$ is ``large'', and  the mapping
$\alpha: \iM \to \iN$  sends the state $\vfi$ of the system of interest
to the state $\vfi \circ \alpha$ at our disposal, then loss of information
takes place and the problem of statistical inference is to reconstruct the
real state from partial information. In this paper we mostly consider
parametric statistical models, a parametric family $\iS:=\{\vfi_\theta:
\theta \in \Theta \}$ of states are given and on the basis of the partial
information the correct value of the parameter should be decided. If the
partial information is the outcome of a measurement, then we have statistical
inference in the very strong sense. However, there are ``more'' quantum
situations, to decide between quantum states on the basis of quantum data,
see Example \ref{E:bipartite} below. The problem we discuss is not the 
procedure of the 
decision about the true state of the system but we want to describe
the circumstances under which this is perfectly possible.

The paper is organized as follows. In the rest of this section we summarize
the relevant basic concepts both in classical statistics in the 
non-commutative framework. Section 2 is about sufficient subalgebras, or
subsystems of a quantum system. Most of the result of this section has been
known but we give a complete presentation and in our proof the operator 
algebraic methods are minimized. Section 3 is devoted to sufficient 
coarse-grainings. The importance of the multiplicative domain of a
completely positive mapping is emphasized here. The factorization theorem of
Section 4 is the main result of the paper. Section 5 connects the exponential
families of the quantum setting to sufficiency problem. In Section 6 the 
equality case in the strong subadditivity of the von Neumann entropy is 
discussed in a possibly infinite dimensional framework and the factorization
result is applied.


In this paper C*-algebras always have a unit $I$. Given a C*-algebra $\iM$,
a state $\vfi$ of $\iM$ is a linear function $\iM \to \bbbc$ such that
$\vfi(I)=1=\|\vfi\|$. (Note that the second condition is equivalent to
the positivity of $\vfi$.) The books \cite{BEH, BR} -- among many others --
explain the basic facts about C*-algebras. The class of finite dimensional
full matrix algebras form a small and algebraically rather trivial subclass
of C*-algebras, but from the view-point of non-commutative statistics,
almost all ideas and concepts appear in this setting. A matrix algebra
$M_n(\bbbc)$ admits a canonical trace $\Tr$ and all states are described
by their densities with respect to $\Tr$. The correspondence is given
by $\vfi(A)=\Tr D_\vfi A\quad$ ($A \in M_n(\bbbc)$) and we can simply identify
the functional $\vfi$ by the density $D_\vfi$. Note that the density is
a positive (semi-definite) matrix of trace 1.

Let $\iM$ and $\iN$ be C*-algebras. Recall that {\bf 2-positivity} of
$\alpha: \iM \to \iN$ means that
$$
\left[ \begin{array}{cc}
\alpha(A)&\alpha(B)\\ \alpha(C)&\alpha(D)\end{array} \right] \geq 0
\qquad \hbox{\ if }\qquad
\left[ \begin{array}{cc}
A& B\\ C& D \end{array} \right] \geq 0\,
$$
for $2\times 2$ matrices with operator entries. It is well-known that
a 2-positive unit-preserving mapping $\alpha$ satisfies the {\bf Schwarz
inequality}
\begin{equation}\label{E:schwarz}
\alpha(A^*A)\geq \alpha(A)^*\alpha(A).
\end{equation}

A 2-positive unital mapping between C*-algebras will be {\bf coarse-graining}.
Here are two fundamental examples.

\begin{ex}
Let $\iX$ be a finite set and $\iN$ be a C*-algebra. Assume that for
each $x \in \iX$ a positive operator $E(x)\in \iN$ is given and
$\sum_x E(x)=I$. In quantum mechanics such a setting is a model for
a measurement with values in $\iX$.

The space $C(\iX)$ of function on $\iX$ is a C*-algebra and the partition
of unity $E$ induces a coarse-graining $\alpha: C(\iX)\to \iN$ given
by $\alpha(f)=\sum_x f(x)E(x)$. Therefore a coarse-graining defined
on a commutative algebra is an equivalent way to give a measurement.
(Note that the condition of 2-positivity is automatically fulfilled on a
commutative algebra.)\qed
\end{ex}

\begin{ex}
Let $\iM$ be the algebra of all bounded operators acting on a Hilbert space
$\iH$ and let $\iN$ be the infinite tensor product $\iM \otimes \iM \otimes
\dots $. (To understand the essence of the example one does not need the
very formal definition of the infinite tensor product.) If $\gamma$
denotes the right shift on $\iN$, then we can define a sequence $\alpha_n$
of coarse-grainings $\iM \to \iN$:
$$
\alpha_n(A):= \frac{1}{n}\big(A+\gamma(A)+\dots +\gamma^{n-1}(A)\big).
$$
$\alpha_n$ is the quantum analogue of the {\bf sample mean}. \qed
\end{ex}

Let $(X_i,\Ae_i,\mu_i)$ be a measure space ($i=1,2$).
Recall that a positive linear map $M:\ L^{\infty}(X_1,\Ae_1,\mu_1)\to
L^{\infty}(X_2,\Ae_2,\mu_2)$ is called a {\bf Markov operator} if  it
satisfies $M1=1$ and $f_n\searrow 0$ implies $Mf_n\searrow 0$.
 For mappings defined between von Neumann algebras, the monotone continuity
is called {\bf normality}.  In case that $\iM$ and $\iN$ are von Neumann
algebras, a coarse-graining $\iM \to \iN$ will be always supposed to be normal.
Our concept of coarse-graining is the analogue of the Markov operator.

We mostly mean that a coarse-graining  transforms observables to observables
corresponding to the {\bf Heisenberg picture} and in this case we assume that
it is unit preserving. The dual of such a mapping acts on states or on density
matrices and it will be called coarse-graining as well.


We recall  some well-known results from  mathematical statistics, see
\cite{Strasser} for details.

Let $(X,{\mathcal  A})$ be a measurable space and let
${\iP }=\{P_{\theta}:\ \theta\in \Theta\}$ be a
set of probability measures on $(X,\Ae)$.
A  sub-$\sigma$-algebra $\mathcal A_0\subset \mathcal A$ is
{\bf sufficient } for
$\iP $ if for all $A\in \mathcal A$, there is an
$\mathcal A_0$-measurable function $f_A$ such that
for all $\theta$,
$$
f_A=P_{\theta}(A|\mathcal A_0)\quad P_{\theta}-\mbox{almost everywhere},
$$
that is,
\begin{equation}
P_{\theta}(A\cap A_0)=\int_{A_0}f_AdP_{\theta}
\end{equation}
for all $A_0\in \mathcal A_0$ and for all $\theta$. It is clear from this
definition that if $\A0$ is sufficient then for all $P_{\theta}$
there is a common version of the
conditional expectation $E_{\theta}[g|\A0]$ for any measurable step function
$g$, or, more generally, for any function $g\in \cap_{\theta\in \Theta}L^1
(X,\Ae,P_{\theta})$.

In the most important case, the family $\iP $ is {\bf dominated}, that is
there is a $\sigma$-finite measure $\mu$ such that $\iP \abs \mu$. The
following lemma is a useful tool in examining sufficiency.

\begin{lemma} \label{lemma:HS}
If $\Pe$ is dominated, then there is a
countable subset $\{P_1,P_2.\dots\}\subseteq \Pe$ such that
$P_{\theta}(A)=0$ holds for all $\theta\in \Theta$ if and only if
$P_n(A)=0$ holds for all $n\in \bbbn$.
\end{lemma}

It follows that if $\iP $ is dominated then there is a (possibly infinite) 
convex combination $P_0=\sum_n c_nP_n$, $P_n\in\Pe$, such that 
$\Pe\equiv P_0$.

For our purposes, it is more suitable to use the following characterization of
sufficiency in terms of randomization.

Let $\Pe_i =\{P_{i,\theta}:\theta\in \Theta\}$ be dominated families
of probability
measures on $(X_i,\Ae_i)$, such that $\Pe_i\equiv\mu_i$, $i=1,2$. We say that
$(X_2,\Ae_2,\Pe_2)$ is a {\bf randomization} of $(X_1,\Ae_1,\Pe_1)$, if there
exists a Markov operator
$M:\ L^{\infty}(X_2,\Ae_2,\mu_2) \to L^{\infty}(X_1,\Ae_1,\mu_1)$,
satisfying
$$
\int (Mf)dP_{\theta,1}=\int fdP_{\theta,2} \qquad (\theta\in\Theta,\
f\in L^{\infty}(X_2,\Ae_2,\Pe_2)).
$$
If also $(X_1,\Ae_1,\Pe_1)$ is a randomization of  $(X_2,\Ae_2,\Pe_2)$, then
$(X_1,\Ae_1,\Pe_1)$ and $(X_2,\Ae_2,\Pe_2)$ are {\bf stochastically 
equivalent}.

For example, let $\Pe\equiv P_0$ and let $\A0\subseteq \Ae$ be a subalgebra.
Then $(X,\A0,\Pe|{\A0})$ is obviously a randomization of $(X,\Ae,\Pe)$, where
the Markov operator is the inclusion
$L^{\infty}(X,\A0,P_0|{\A0})\to L^{\infty}(X,\Ae,P_0)$. On the
other hand, if $\A0$ is sufficient, then the map
$$
f\mapsto E[f|\A0],\qquad E[f|\A0]=E_{\theta}[f|\A0],\quad P_{\theta}-
\mbox{almost everywhere},
$$
is a Markov
operator $L^{\infty}(X,\Ae,P_0)\to L^{\infty}(X,\A0,P_0|_{\A0})$ and
$$
\int E[f|\A0]dP_{\theta}|_{\A0}=\int fdP_{\theta}
\qquad (f\in L^{\infty}(X,\Ae,P_0),
\ \theta\in\Theta).
$$
We have the following characterizations of sufficient subalgebras.

\begin{prop} Let $\Pe$ be a dominated family and let $\A0\subseteq \Ae$
be a sub-$\sigma$-algebra. The following are equivalent.
\begin{enumerate}
\item[(i)] $\A0$ is sufficient for $\Pe$
\item[(ii)] There exists a measure $P_0$ such that $P_0\equiv\Pe$ and
${dP_{\theta}/}{dP_0}$ is $\A0$-measurable for all $\theta$.
\item[(iii)] $(X,\Ae,\Pe)$ and  $(X,\A0,\Pe|{\A0})$ are
stochastically equivalent
\end{enumerate}
\end{prop}

It follows that
if $\Pe\equiv\mu$, then the sub-$\sigma$-algebra generated
by the functions $\{{dP_{\theta}}/{d\mu}: \theta\in \Theta\}$  is
sufficient for $\Pe$, moreover, it is contained in any other
sufficient subalgebra in $\Ae$. Such subalgebras are called
{\bf minimal sufficient}.

Next we formulate a non-commutative setting. Let $\iM$ be a von Neummann
algebra and $\iM_0$ be its von Neumann subalgebra. Assume that a family
$\iS:=\{ \vfi_\theta: \theta \in \Theta\}$ of normal states are given.
$(\iM,\iS)$ is called {\bf statistical experiment}. The subalgebra $\iM_0
\subset\iM$ is {\bf sufficient} for  $(\iM,\iS)$ if for every $a \in \iM$,
there is $\aa(a) \in \iM_0$ such that
\begin{equation}
\vfi_\theta(a)=\vfi_\theta(\aa(a))  \qquad (\theta \in \Theta)
\end{equation}
and the correspondence $a \mapsto \alpha(a)$ is a coarse-graining.
(Note that a positive mapping is automatically completely
positive if it is defined on a commutative algebra.)

\begin{ex}\label{E:bipartite}
Consider a bipartite system $\iH=\iH_A \otimes \iH_B$ and a family
$\{ \vfi_\theta:  \theta \in \Theta\}$ of states on $\iH$. Assume that
the expectation value of all observables localized at $A$ is known to
us, that is, we know the restriction of $\vfi_\theta$'s to $B(\iH_A)$
(or the reduced density matrices). This information is not sufficient
in general to decide about $\theta$. We impose the further condition
that $\iH_A=\iH_L\otimes \iH_R$ and the factorization
$$
\vfi_\theta = \vfi_\theta^0 \otimes \vfi_{RB},
$$
where $\vfi_\theta^0$ is a state on $B(\iH_L)$ and the state
$\vfi_{RB}$ of  $B(\iH_R)\otimes B(\iH_B)$ is independent of the parameter
$\theta$. In this case the restriction of the unknown state to $B(\iH_L)$
determines the true value of the parameter $\theta$ and $\vfi_\theta$
is recovered uniquely.

The subalgebra $B(\iH_L)$ is sufficient and the example is close to typical. 
In the general case, however, the relation of the subalgebras $B(\iH_L)$ and
$B(\iH_A)$ is more subtle. \qed
\end{ex}

The following lemma is a quantum version of Lemma \ref{lemma:HS}.

\begin{lemma}\label{lemma:qhs}
Assume that the von Neumann algebra $\iM$ admits a faithful normal state
$\psi$.
Let $\iS=\{\vfi_{\theta}:\ \theta\in\Theta\}$ be a family of normal
states on $\iM$. Then there is a sequence $(\vfi_n)$
of states in $\iS$ and a normal state
$$
\omega = \sum_{n=1}^{\infty} \lambda_n \vfi_n
$$
such that $\supp\vfi_{\theta}\leq\supp\omega$ for all $\theta\in \Theta$.
\end{lemma}

{\it Proof:} Let $\{ p_i:\ i\in I\}$ be a set of pairwise orthogonal
projections in $\iM$,  then $\psi(p_i)>0$ and $\psi(\sum_i p_i)\leq 1$,
therefore any such set  must be at most countable.

We set
$$
\Pe=\{ p_\theta:=\supp \vfi_\theta:\ \theta\in\Theta\}
$$
and show that there is a countable subset
$\{p_1,p_2,\dots\}\subset \Pe$, such that
$\sup_\theta p_\theta= \sup_n p_n$

Let $\mathcal C$ be a set of at most countable subsets in $\Pe$, ordered by
inclusion. Consider all chains in $\mathcal C$, such that if $C\subset D$ in
the chain, then $\sup  C \neq \sup D$. It is clear that each such chain
has at most countably many elements. Let $\{ C_1,C_2,\dots\}$ be a
maximal such chain and let  $C=\cup_n C_n=\{p_1,p_2,\dots\}$. Then
$\sup_n p_n=\sup_\theta p_\theta$. Indeed, if $\sup_n p_n\neq \sup_\theta
p_\theta$, then there is an element $p\in \Pe$, such that $\sup C\neq \sup
C\cup \{p\}$, which contradicts the maximality of $\{C_1,C_2,\dots\}$.

Let now $\vfi_1,\vfi_2,\dots$ be elements is $\iS$ such that $\supp\vfi_n=p_n$.
Choose a sequence $\lambda_1,\lambda_2,\dots$ such that
$\lambda_n>0$ for all $n$ and $\sum_n\lambda_n=1$ and put
$\omega =\sum\lambda_n\vfi_n$. Then it is clear that $\supp\omega=\sup_n p_n$
and $\supp\vfi_\theta\leq \supp\omega$ for all $\theta$.
\qed

Throughout the paper, we  suppose that the hypothesis of the above lemma is
satisfied, that is, the von Neumann algebras considered admit a faithful
normal state. The algebra $B(\iH)$ satisfies this condition if and only if
the Hilbert space $\iH$ is separable.

When the states $\vfi_n$ belong to $\iS$ and for
$$
\omega := \sum_{n=1}^{\infty} \lambda_n \vfi_n
$$
the condition $\supp\vfi_{\theta}\leq\supp\omega$ holds for all 
$\theta\in \Theta$, we say that $\iS$ {\bf is dominated by} $\omega$.

\section{Sufficient subalgebras}

In the study of sufficient subalgebras monotone quasi-entropy quantities
could be useful. The {\bf relative entropy} and the {\bf transition 
probability} are examples of those \cite{petz1986b, OP}.

Let $\varphi$ and $\omega$ be normal states of a von Neumann algebra and
let $\xi_\varphi$ and $\xi_\omega$ be the representing vectors of these states
from the natural positive cone. Then the {\bf transition probability} is defined as
$$
P_A(\varphi, \omega)=\langle \xi_\varphi,\xi_\omega \rangle.
$$
In case of density matrices this reduces to $P_A(D_1, D_2) = \Tr (D^{1/2}_1 D^{1/2}_2)$.

\begin{thm}\label{thm:1}
Let $(\iM, \{\vfi_\theta:\theta \in \Theta\} )$ be a statistical experiment
and let $\iM_0 \subset \iM$ be von Neumann algebras. Assume that 
$\{\vfi_\theta:\theta \in \Theta\}$ is dominated by a faithful normal
state $\omega$. Then the following conditions are equivalent.

\begin{enumerate}
\item[(i)]  $\iM_0$ is sufficient for $(\vfi_\theta )$.
\item[(ii)] $P_A(\vfi_\theta ,\omega )=P_A(\vfi_\theta|\iM_0 ,\omega|\iM_0 )$ for
all $\theta$.
\item[(iii)]  $[D\vfi_\theta , D\omega ]_t=[D(\vfi_\theta|\iM_0), D(\omega
|\iM_0)]_t\,$ for every real $t$ and for every $\theta$.
\item[(iv)] $[D\vfi_\theta,D\omega]_t\in \iM_0$ for all real $t$ and every $\theta$.
\item[(v)]  The generalized conditional expectation
$E_\omega :\iM \to \iM_0$ leaves all the states $\vfi_\theta$ invariant.
\end{enumerate}
\end{thm}

Note that condition (iii) is formulated in terms of Connes' Radon-Nikodym cocycle
and the generalized conditional expectation  appearing in (iv) is discussed in the appendix.

The theorem is essentially Thm 9.5 from \cite{OP} and we give the detailed proof in
the finite dimensional situation. The following two lemmas will be used.

\begin{lemma}\label{lemma:monot}
Let $T:B(\iH)\to B(\iK)$ be a coarse-graining sending density matrices
to densities. Let $D_1$ and $D_2$ be density matrices acting on the
Hilbert space $\iH$. Then
$$
P_A(D_1, D_2) \leq P_A(T(D_1), T(D_2))
$$
\end{lemma}

{\it Proof:}
On the Hilbert space $B(\iH)$ one can define an operator $\Delta$ as
$$
\Delta a = D_2 a D_1^{-1} \qquad (a \in B(\iH)),
$$
where the generalized inverse $ D_1^{-1}$ is determined by the
relation $D_1 D_1^{-1}= D_1^{-1}D_1=\supp D_1$. This is the
so-called {\bf relative modular operator} and it is the product of two
commuting positive operators: $\Delta=LR$, where
$$
La= D_2a\quad {\rm and}\quad Ra=a D_1^{-1}  \qquad (a \in B(\iH)).
$$

We have
$$
P_A(D_1, D_2) = \<D^{1/2}_1, \Delta^{1/2} D^{1/2}_1\>.
$$

Set
$$
\Delta a = D_2 a D_1^{-1} \quad (a \in B(\iH))\quad\hbox{and}\quad
\Delta_0 x = T(D_2) x T(D_1)^{-1} \quad (x \in B(\iK)).
$$
$\Delta$ and $\Delta_0$ are operators on the spaces $B(\iH)$ and
$B(\iK)$. (They become  Hilbert space with the
Hilbert-Schmidt inner product.) The transition probabilities
are expressed by the resolvent of relative modular operators:
\begin{eqnarray*}
P_A(D_1, D_2)& =& \<D^{1/2}_1, \Delta^{1/2} D^{1/2}_1\> \\
&=& \frac{1}{\pi}\int^{\infty}_{0} t^{-1/2}- t^{1/2} \<D^{1/2}_1,
(\Delta + t)^{-1} D^{1/2}_1\> \,dt \\
P_A(T(D_1), T(D_2))& =& \<T(D_1)^{1/2}, \Delta_0^{1/2} T(D_1)^{1/2}\> \\
&=& \frac{1}{\pi}\int^{\infty}_{0} t^{-1/2}- t^{1/2} \<T(D_1)^{1/2},
(\Delta_0 + t)^{-1} T(D_1)^{1/2}\> \,dt
\end{eqnarray*}
where the identity
$$
x^{1/2}= \frac{1}{\pi}\int^{\infty}_{0} t^{-1/2}- t^{1/2}(x+t)^{-1}\,dt
$$
is used. Let us define the operator
\begin{equation}
V(xT(D_1)^{1/2}+\xi)=T^*(x)D_1^{1/2}
\end{equation}
where $\xi\in [\iB(\iK)T(D_1)^{1/2}]^{\perp}$. Then $V$
is a contraction:
\begin{eqnarray*}
\Vert T^*(x)D_1^{1/2}\Vert^2 &=& \Tr D_1 T^*(x^*)T^*(x)\leq
\Tr D_1 T^*(x^* x)=\Tr T(D_1) x^* x=\\
&=& \Vert xT(D_1)^{1/2}\Vert^2\leq \Vert xT(D_1)^{1/2}+\xi\Vert^2
\end{eqnarray*}
since the Schwarz inequality is applicable to $T^*$.
Let now $p_1=\supp D_1$ and  $q_1=\supp T(D_1)$. Since $T^*$ is unital,
$0=\Tr T(D_1)(1-q_1)=\Tr D_1 (1-T^*(q_1))$ and therefore $p_1\leq T^*(q_1)$.
The Schwarz inequality (\ref{E:schwarz}) now implies
\begin{eqnarray*}
\< V(xT(D_1)^{1/2}+\xi),\Delta V(xT(D_1)^{1/2}+\xi)\>&=&
\Tr  D_2T^*(x)p_1T^*(x^*)\leq \Tr D_2T^*(xq_1x^*)\\ =
\< xT(D_1)^{1/2},\Delta_0 xT(D_1)^{1/2}\>&\leq&
\< xT(D_1)^{1/2}+\xi,\Delta_0(xT(D_1)^{1/2}+\xi)\>
\end{eqnarray*}
where the last inequality follows from
$$
\<\Delta_0xT(D1)^{1/2},\xi\>=\<T(D_2)xT(D_1)^{-1}T(D_1)^{1/2},\xi\>=0
$$
It follows that
\begin{equation}
V^*\Delta V \leq \Delta_0\,.
\end{equation}
The function $y \mapsto (y+t)^{-1}$ is operator monotone (decreasing) and
operator convex, hence
\begin{equation}\label{E:op}
(\Delta_0+t)^{-1}\leq (V^* \Delta V +t)^{-1}\leq V^*(\Delta +t)^{-1}V
\end{equation}
(see \cite{HanPed}). Since $VT(D_1)^{1/2}= D_1^{1/2}$,  this implies
\begin{equation}
\<D^{1/2}_1, (\Delta + t)^{-1} D^{1/2}_1\> \geq
\<T(D_1)^{1/2}, (\Delta_0 + t)^{-1} T(D_1)^{1/2}\,.
\end{equation}
By integrating this inequality we have the monotonicity theorem
from the above integral formulas. \qed

Condition (i) implies that $P_A(\vfi_\theta ,\omega )=P_A(\vfi_\theta|\iM_0,
\omega |\iM_0 )$ due to the monotonicity of the transition probability under
completely positive mappings. Indeed, if $\alpha$ leaves all $\varphi_\theta$
invariant, then $\omega \circ \alpha=\omega$.

Now we are in the position to analyze the case of equality.

\begin{lemma}\label{lemma:equality}
If
$$
P_A(D_1,D_2)= P_A(T(D_1),T(D_2)),
$$
then
$$
T^*(T(D_2)^{it}T(D_1)^{-it})p_1=D_2^{it}D_1^{-it}p_1\, ,
$$
where $p_1=\supp D_1$.
\end{lemma}

{\it Proof:}
From the integral formula for the transition probability we have
\begin{equation}
\<T(D_1)^{1/2}, V^*(\Delta + t)^{-1}V T(D_1)^{1/2}\> =
\<T(D_1)^{1/2}, (\Delta_0 + t)^{-1} T(D_1)^{1/2}\>\,.
\end{equation}
for all $t>0$. This equality together with the operator inequality
(\ref{E:op}) gives
\begin{equation}
V^*(\Delta + t)^{-1} D_1^{1/2}=(\Delta_0 + t)^{-1} T(D_1)^{1/2}
\end{equation}
for all $t>0$. Differentiating by $t$ we have
\begin{equation}
V^*(\Delta + t)^{-2} D_1^{1/2}=(\Delta_0 + t)^{-2} T(D_1)^{1/2}
\end{equation}
and we infer
\begin{eqnarray*}
\Vert V^*(\Delta + t)^{-1} D_1^{1/2}\Vert^2
&=&\<(\Delta_0 + t)^{-2} T(D_1)^{1/2}, T(D_1)^{1/2}\>\\
&=& \<V^*(\Delta + t)^{-2} D_1^{1/2},T(D_1)^{1/2}\>\\
&=& \Vert (\Delta + t)^{-1} D_1^{1/2}\Vert^2
\end{eqnarray*}
When $\Vert V^*\xi\Vert = \Vert \xi \Vert$ holds for a contraction $V$,
it follows that $VV^*\xi=\xi$. In the light of this remark we arrive at
the condition
$$
VV^*(\Delta + t)^{-1} D_1^{1/2}=(\Delta + t)^{-1} D_1^{1/2}
$$
and
\begin{eqnarray*}
V(\Delta_0 + t)^{-1} T(D_1)^{1/2}
& =& VV^*(\Delta + t)^{-1} D_1^{1/2}\\
& = & (\Delta + t)^{-1} D_1^{1/2}
\end{eqnarray*}
By Stone-Weierstrass approximation we have
\begin{equation}
V f(\Delta_0) T(D_1)^{1/2}=f(\Delta) D_1^{1/2}
\end{equation}
for continuous functions. In particular for $f(x)=x^{i t}$ we
have
\begin{equation}\label{E:ns}
T^*\big(T(D_2)^{it}T(D_1)^{-it}\big)p_1= D_2^{it} D_1^{-it}p_1\,.
\end{equation}
This condition is necessary and sufficient for the equality. \qed

The previous lemma shows that condition (ii) implies (iii) and it  is clear that (iii) implies (iv). We prove that
(iv) implies (i).

Let $\iM_1$ be the subalgebra generated by
$\{ [D\vfi_\theta,D\omega]_t,\ t\in\ \mathbb{R}\}$ and let $\omega_1$, $\vfi_1$ be the restrictions of $\omega$,
$\vfi_\theta$ to $\iM_1$. Then $[D\vfi_\theta,D_\omega]_t$
satisfies the cocycle condition for $\sigma_t^{\omega_1}$ and therefore there  is a weight $\psi$ on $\iM_1$,
such that $[D\psi,D\omega_1]_t=[D\vfi_\theta,D\omega]_t$.

On the other hand, $\iM_1$ is invariant
under the modular group $\sig_\omega^t$, hence there exists a conditional expectation
$F:\ \iM\to\iM_1$ preserving $\omega$ and
$$
[D\psi\circ F,D\omega]_t=[D\psi,D\omega_1]_t=[D\vfi_\theta,D\omega]_t,\quad \forall t
$$
It follows that $\psi\circ F=\vfi_\theta$, therefore   $\psi=\vfi_1$ and $F$ preserves also $\vfi_\theta$.

Let now (iv) be satisfied, then $\iM_1\subseteq \iM_0$.  The conditional expectation $F$ is a coarse-graining
$\iM\to\iM_0$ preserving all $\vfi_\theta$ and (i) follows.

Next, we want to show that (iii) implies (v).
Let $E: \iM\to \iM_0$ be the trace preserving conditional expectation.
Then the generalized conditional expectation $E_\omega:\iM\to \iM_0$ acts as
$$
E_\omega(a)=E(D)^{-1/2}E(D^{1/2}aD^{1/2})E(D)^{-1/2}
$$
We have to show that
$$
\Tr E(D_\theta) E_\omega(a)=\Tr D_\theta a
$$
which is equivalently written as
$$
\Tr E(D_\theta)^{1/2}E(D)^{-1/2}E(D^{1/2}aD^{1/2})E(D)^{-1/2} E(D_\theta)^{1/2}
=\Tr D_\theta a
$$
By analytic continuation from condition (iii), we have
$$
E(D_{\theta})^{1/2}E(D)^{-1/2}=D_{\theta}^{1/2}D^{-1/2}
$$
It follows that
\begin{eqnarray*}
\Tr E(D_\theta) E_\omega(a)&=&\Tr E(D_\theta)^{1/2}E(D)^{-1/2}D^{1/2}aD^{1/2}
E(D)^{-1/2} E(D_\theta)^{1/2}=\\
&=&\Tr D_{\theta}D^{-1/2}D^{1/2}aD^{1/2}D^{-1/2}D_{\theta}^{1/2}=
\Tr D_{\theta}a\,.
\end{eqnarray*}

The implication (v) $\to$ (i) is trivial. \qed

\section{Sufficient statistic and coarse-graining}

A classical {\bf sufficient statistic} for the family $\iP$ is a
 measurable mapping $T:\ (X,\mathcal A)\to (X_1,\mathcal A_1)$
such that  the generated sub-$\sigma$-algebra $T^{-1}(\mathcal A_1)\subset
\mathcal A$ is sufficient for $\iP $. To any statistic $T$, we associate a
Markov operator
$$
\tilde T:\ L^{\infty}(X_1,\Ae_1,P_0^T)\to L^{\infty}(X,\Ae,P_0),\quad
(\tilde Tg)(x)=g(T(x))
$$
Obviously,  $(X_1,\Ae_1,\Pe^T)$ is a randomization of $(X,\Ae,\Pe)$.
As in the case of subalgebras, we have

\begin{prop}
The statistic $T:\ (X,\Ae)\to (X_1,\Ae_1)$ is sufficient for
 $\Pe$ if and only if
$(X,\Ae,\Pe)$ and $(X_1,\Ae_1,\Pe^T)$ are stochastically equivalent.
\end{prop}

\begin{prop}(Factorization criterion)
Let $\Pe \abs \mu$. The statistic
$T:\ (X,\Ae)\to(X_1,\Ae_1)$ is sufficient for $\Pe$  if and only if
there is an $\Ae_1$-measurable function $g_{\theta}$ for all $\theta$ and an
$\Ae$-measurable function $h$ such that
$$
\frac{dP_{\theta}}{d\mu}(x)=g_{\theta}(T(x))h(x)\quad P_{\theta}-\mbox{almost
everywhere}
$$
\end{prop}

Let $\iN$, $\iM$ be C*-algebras and let $\sigma:\ \iN\to \iM$ be a
coarse-graining. We say that $\sigma$ is sufficient for the statistical
experiment $(\iM,\vfi_\theta)$ if there exists  a
coarse-graining $\beta:\iM \to \iN$ such that
$\vfi_\theta \circ \sig \circ \beta = \vfi_\theta$ for every $\theta$.

Let $\omega=\sum_n\lambda_n\vfi_n$
be the normal state obtained in Lemma \ref{lemma:qhs} and let $p=\supp \omega$,
$q=\supp \omega\circ\sig$. Let us define the map $\alpha:\ q\iN q\to p\iM p$
by $\alpha(a)=p\sig(a)p$, then $\alpha$ is a coarse-graining such that
$\vfi_\theta\circ \sig(a)=\vfi_\theta\circ\alpha(qaq)$ for all $\theta$ and
$\alpha^*_\omega=\sig^*_\omega$, where the dual $\sig^*_\omega$ is defined in
the Appendix. We check that $\alpha$ is sufficient for
$(p\iM p,\vfi_\theta|_{p\iM p})$ if and only if $\sigma$ is sufficient for
$(\iM,\vfi_\theta)$. Indeed, let $\tilde\beta:p\iM p\to q\iN q$ be a
coarse-graining such that
$\vfi_\theta|_{p\iM p}\circ\alpha\circ\tilde\beta=
\vfi_\theta|_{p\iM p}$ and
let $\beta:\iM\to \iN$ be defined by
$$
\beta(a)=\tilde\beta(pap)+\omega(a)(1-q)
$$
Then $\beta$ is a coarse-graining and
$$
\vfi_\theta\circ\sig\circ\beta(a)=\vfi_\theta\circ\sig(q\beta(a)q))=
\vfi_\theta\circ\alpha\circ\tilde\beta(pap)=\vfi_\theta(pap)=\vfi_\theta(a)
$$
The converse is proved similarly, taking $\tilde\beta(a)=q\beta(a)q$ for
$a\in p\iM p$. Therefore  we may, and will, suppose that both $\omega$
and $\omega\circ\sig$ are faithful.

Let us recall the following property of coarse-grainings.

\begin{lemma}\label{lemma:Nsig}
Let $\iM$ and $\iN$ be C*-algebras and let $\sigma:\iN \to \iM$ be a
coarse-graining. Then
\begin{equation}
\iN_\sigma:=\{a \in \iN: \sig(a^*a)=\sig(a)\sig(a)^*\mbox{\ and\ }
\sig(aa^*)=\sig(a)^*\sig(a)\}
\end{equation}
is a subalgebra of $\iN$ and
\begin{equation}
\sig(ab)=\sig(a)\sig(b) \quad \mbox{and}\quad \sig(ba)=\sig(b)\sig(a)
\end{equation}
holds for all $a \in \iN_\sigma$ and $b \in \iN$.
\end{lemma}

{\it Proof:}
The proof is based only on the Schwarz inequality
$$
\sigma(x^*x)\geq \sigma(x)^*\sigma(x).
$$
From this we have
\begin{eqnarray*}
t\big(\sigma(a)\sigma(b)+ \sigma(b)^*\sigma(a)^*\big)
& =&\sigma(ta^*+b)^*\sigma(ta^*+b)-t^2\sigma(a)\sigma(a)^*-\sigma(b)^*\sigma(b)
\\ & \leq & \sigma(ta^*+b)^*\sigma(ta^*+b)-t^2\sigma(aa^*)-\sigma(b)^*\sigma(b)\\
& = & t\sigma(ab+b^*a^*)+\sigma(b^*b)-\sigma(b)^*\sigma(b)
\end{eqnarray*}
for a real $t$ and $a \in \iN_\sigma$.
Divide the inequality by $t$ and let $t \to \pm \infty$. Then
$$
\sigma(a)\sigma(b)+ \sigma(b)^*\sigma(a^*)=\sigma(ab+b^*a^*)
$$
and similarly
$$
\sigma(a)\sigma(b)- \sigma(b)^*\sigma(a)^*=\sigma(ab-b^*a^*).
$$
Adding these two inequalities we have
$$
\sigma(ab)=\sigma(a)\sigma(b).
$$
\qed

We call the subalgebra $\iN_\sigma$ the {\bf multiplicative domain} 
of $\sigma$.

Let now $\iN$ and $\iM$ be von Neumann algebras and let $\omega$ be a faithful
normal state on $\iM$ such that $\omega\circ\sig$ is also faithful. Let
$$
\iN_1=\{ a\in \iN,\ \sigma_\omega^*\circ\sig(a)=a\}
$$
It was proved in \cite{petz1988} that $\iN_1$ is a subalgebra of $\iN_\sig$,
moreover, $a\in\iN_1$ if and only if $\sig(a^*a)=\sig(a)^*\sig(a)$ and
$\sig(\sig_t^{\omega\circ\sig}(a))=\sig_t^{\omega}(\sig(a))$. The restriction
of $\sig$ to $\iN_1$ is an isomorphism onto
$$
\iM_1=\{b\in\iM,\ \sig\circ\sig_\omega^*(b)=b\}
$$

The following Theorem was proved in \cite{petz1988} in the case when
$\vfi_\theta$ are  faithful states.

\begin{thm}\label{thm:3}
Let $\iM$ and $\iN$ be von Neumann algebras and let $\sig:\iN \to \iM$ 
be a coarse-graining. Suppose that $(\iM,\vfi_\theta )$ is a statistical 
experiment dominated by a state $\omega$ such that both $\omega$ and 
$\omega\circ\sig$ are faithful and normal.

Then following properties are equivalent:
\begin{enumerate}
\item[(i)] $\sig(\iN_\sig)$ is a sufficient subalgebra
for $(\iM,\vfi_\theta )$.
\item[(ii)]  $\sig$ is a sufficient coarse-graining for $(\iM,\vfi_\theta)$.
\item[(iii)] $P_A(\vfi_\theta,\omega)=
P_A(\vfi_\theta\circ\sig,\omega\circ\sig)$
\item[(iv)] $\sig([D\vfi_\theta\circ\sig,D\omega\circ\sig]_t)=
[D\vfi_\theta,D\omega]_t$
\item[(v)]  $\iM_1$
is a sufficient subalgebra for $(\iM,\vfi_\theta)$.
\item[(vi)] $\vfi_\theta\circ\sig\circ\sig_\omega^*=\vfi_\theta$.
\end{enumerate}
\end{thm}

{\it Proof.} Suppose (i), then there is a coarse-graining $\gamma:\iM \to
\sig(\iN_\sig)$, preserving $\vfi_\theta$. It is easy to see
that the restriction of $\sig$  to $\iN_\sig$ is invertible.
Let $\alpha$ be the inverse of this restriction and put
$$
\beta=\alpha\circ\gamma
$$
Then $\beta: \iM \to \iN $ is a coarse-graining such that
$\vfi_\theta\circ\sig\circ\beta=\vfi_\theta$   and  (ii) is proved.

The implications (ii) $\to$ (iii) and (iii) $\to$ (iv) follow from Lemmas
\ref{lemma:monot} and \ref{lemma:equality}.

Suppose (iv) and denote $u_t=[D\vfi_\theta\circ\sig,D\omega\circ\sig]_t$,
$v_t=[D\vfi_\theta,D\omega]_t$. Then we have $\sig(u_t)=v_t$ for all
$t$. Let $p_\theta=\supp\vfi_\theta$, $q_\theta=\supp\vfi_\theta\circ\sig$.
Putting $t=0$ in the condition (iv), we get $\sig(q_\theta)=p_\theta$ and
$$
\sig(u_tu_t^*)=\sig(q_\theta)=p_\theta=v_tv_t^*=\sig(u_t)\sig(u_t)^*
$$
On the other hand, $\sig(u_t)^*\sig(u_t)\leq \sig(u_t^*u_t)$ by Schwartz
inequality and from
\begin{eqnarray*}
\omega(\sig(u_t)^*\sig(u_t))&=&\omega(v_t^*v_t)=
\omega(\sig_t^{\omega}(p_\theta))=\omega(p_\theta)\\
\omega(\sig(u_t^*u_t))&=&\omega\circ\sig(u_t^*u_t)=
\omega\circ\sig(\sig_t^{\omega\circ\sig}(q_\theta))=\omega(p_\theta)
\end{eqnarray*}
we get $\sig(u_t^*u_t)=\sig(u_t)^*\sig(u_t)$. Hence
$u_t\in \sig(\iN_\sig)$ for all $t$.
Further, by the cocycle condition and Lemma \ref{lemma:Nsig},
$$
\sig(\sig_t^{\omega\circ\sig}(u_t))=\sig(u_s^*u_{s+t})=v_s^*v_{t+s}=
\sig_t^\omega(\sig(u_t))
$$
therefore $v_t\in \iM_1$ and by Theorem \ref{thm:1}, $\iM_1$ is sufficient and
(v) is proved.
As $\iM_1$ is a subalgebra in $\sig(\iN_\sig)$, this implies (i).

Finally, we prove that (ii) is equivalent to (vi).
First, note that a coarse-graining is sufficient for
$(\iM,\vfi_\theta)$ if and only if it is sufficient for
$(\iM,\psi_\theta)$, where
$$
\psi_\theta=\varepsilon \vfi_\theta+(1-\varepsilon)\omega
$$
for some $0<\varepsilon<1$.

As the states $\psi_\theta$ are faithful and
$\omega=\sum_n\lambda_n\psi_n$, it follows from the results in \cite{petz1988}
that  $\sig$ is sufficient if and only if
$\psi_\theta\circ\sig\circ\sig_\omega^*=\psi_\theta$
for all $\theta$.
Since, by definition, $\omega\circ\sig\circ\sig^*_\omega=\omega$, this is
equivalent to (vi).
\qed

 Let $\iM_0\subset \iM$ be a subalgebra. From the above  theorem,
 together with the remarks preceding Lemma \ref{lemma:Nsig},
we have  a generalization of Theorem \ref{thm:1} to the case that $\supp \omega =p$ and $\supp \omega|_{\iM_0}=q$.
Namely, $\iM_0$ is sufficient for $(\iM,\vfi_\theta)$ if and only if the coarse-graining $\alpha:s q\iM_0q\to p\iM p$, $\alpha(qaq)=pap$ is
sufficient for the restricted experiment.

{\it Remark.} Let $S(\vfi,\omega)$ be the relative entropy and suppose that
$S(\vfi_\theta,\omega)$ is finite for all $\theta$. Then the condition (iii)
can be replaced by
$$
S(\vfi_\theta,\omega)=S(\vfi_\theta\circ\sig,\omega\circ\sig)
$$
This can be proved similarly as for the transition probability, using the
formula
\begin{equation}\label{eq:intlog}
\log x = \int_0^\infty (1+t)^{-1}-(x+t)^{-1}dt\, .
\end{equation}
The equality in inequalities for entropy quantities was studied also in
\cite{MBR1}.
\qed

The previous theorem applies to a measurement which is essentially
a positive mapping $\iN \to \iM$ from a commutative algebra. The concept
of sufficient measurement appeared also in \cite{BN-G}. For a non-commuting
family of states, there is no sufficient measurement.

\section{Factorization}\label{sec:factorization}

Let  $\iM$ be a von Neumann  algebra and let $\omega$ be a
faithful state on $\iM$. Let $\iM_0\subset \iM$ be a subalgebra and assume
that  it is invariant under the modular group $\sig_t^\omega$ of $\omega$.
Let $\iM_1=\iM_0'\cap \iM$ be the relative commutant.
We show that $\iM_1$  is invariant under $\sig_t^\omega$ as well.
If $a\in \iM_0$ and $b\in \iM_1$, then for $t\in \mathbb{R}$, we have
$$
a\sigma_t^\omega(b)=\sigma_t^{\omega}\big(\sigma_{-t}^{\omega}(a)b\big)=
\sigma_t^{\omega}\big(b \sigma_{-t}^{\omega}(a)\big)=
\sigma_t^\omega(b)a
$$
Hence $\iM_1$ is invariant under $\sigma_t^\omega$. Let $\omega_0$, $\omega_1$  be the restrictions of $\omega$ to $\iM_0$ and $\iM_1$.
 Then $\sigma_t^\omega|_{\iM_0}=\sigma_t^{\omega_0}$ and $\sig_t^\omega|_{\iM_1}=\sig_1^{\omega_1}$
are known facts in modular theory.

Recall that the {\bf entropy} of a state $\vfi$ of a C*-algebra is defined as
$$
S(\vfi):=\sup \Big\{ \sum_i \lambda_i S(\vfi_i\|\vfi):\sum_i \vfi_i=\vfi\Big\},
$$
see (6.9) in \cite{OP}.
For the sake of simplicity, we will suppose in the rest of this section that
the state $\omega$ has finite von Neumann entropy $S(\omega)$. Then $\iM$
must be a countable direct sum of type I factors, see Theorem 6.10. in \cite{OP}. 
Let $\tau$ be the canonical normal semifinite trace
on $\iM$ and let $D_\omega$ be the  density of $\omega$ with respect to $\tau$, then
$$
\sig_t^\omega(a)=D_\omega^{it}aD_\omega^{-it}, a\in\iM.
$$

As the subalgebras $\iM_0$ and $\iM_1$ are invariant under $\sigma_t^\omega$, we 
have by Proposition 6.7. in \cite{OP}
that $S(\omega_0), S(\omega_1)\le S(\omega)<\infty$. It follows that both $\iM_0$ 
and $\iM_1$ must be
countable direct sums of type I factors as well.

Let $D_{\omega_0}\in \iM_0$ and $D_{\omega_1}\in\iM_1$ be the  densities of 
$\omega_0$ and $\omega_1$ with respect to the canonical traces
$\tau_0:=\tau|\iM_0$ and $\tau_1:=\tau|\iM_1$. Then for $a\in \iM_0$,
$$
D_\omega^{it}aD_\omega^{-it}=
\sig_t^\omega(a)=\sig_t^{\omega_0}(a)=
D_{\omega_0}^{it}aD_{\omega_0}^{-it}.
$$
It follows that $w_t:=D_{\omega_0}^{-it}D_\omega^{it}$ is  a unitary operator 
in $\iM_1$ and
the  operators $D_{\omega_0}^{it}$ and $D_\omega^{is}$ commute for all
$t,s\in\mathbb{R}$. It is easy to see that $w_t$ is a strongly
 continuous one-parameter group. Moreover, we have for $a\in\iM_1$,
$$
w_taw_t^*=D_{\omega}^{it}aD_{\omega}^{-it}=
\sigma_t^{\omega_1}(a)=D_{\omega_1}^{it}aD_{\omega_1}^{-it}
$$
Therefore,  the unitary $z_t=D_{\omega_1}^{-it}w_t$ is  in the center of
$\iM_1$. Again,  $w_t$ and $D_{\omega_1}^{is}$ commute for all $t$, $s$
and it is easy to see that $z_t=z^{it}$ for some positive element 
$z$ in the center of $\iM_1$. Putting all together, we get
\begin{equation}\label{eq:factoromega}
D_\omega=D_{\omega_0}D_{\omega_1}z
\end{equation}

The following theorem is a  generalization  of the classical 
factorization theorem.

\begin{thm}\label{thm:2} 
Let $(\iM,\iS)$ be a statistical experiment dominated by a faithful normal state  
$\omega$ such that $S(\omega)<\infty$. Let  $\iM_0\subset \iM$ be a von Neumann 
subalgebra invariant with respect to the modular group $\sigma_t^\omega$. Then 
$\iM_0$ is sufficient for $\iS$ if and only if
\begin{equation}\label{eq:factortheta}
D_\theta= D_{\theta,0}D_{\omega_1}z,
\end{equation}
where $D_\theta$, $D_{\theta,0}$ and $D_{\omega_1}$ are the densities of 
$\vfi_\theta$, $\vfi_\theta|_{\iM_0}$ and $\omega | \iM_0' \cap \iM$, respectively
and $z$ is a positive operator from the centre of $\iM_0' \cap \iM$.
\end{thm}

{\it Proof.} By the assumptions and (\ref{eq:factoromega}), we have
$D_\omega^{it}=D_{\omega_0}^{it}D_{\omega_1}^{it}z^{it}$.
If $\iM_0$ is sufficient, then
$$
u_t:=D_\theta^{it}D_\omega^{-it}=[D\vfi_\theta,D\omega]_t=[D\vfi_\theta|_{\iM_0},D\omega_0]_t
=D_{\theta,0}^{it}D_{\omega_0}^{-it},
$$
hence $D_\theta^{it}=u_tD_\omega^{it}=D_{\theta,0}^{it}D_{\omega_1}^{it}z^{it}$
and (\ref{eq:factortheta}) follows.

Conversely, let (\ref{eq:factortheta}) be true, then $u_t=D_{\theta,0}^{it}D_{\omega_0}^{-it}$ and $\iM_0$ is sufficient.
\qed

The essence of the factorization (\ref{eq:factoromega}) is that the first factor
depends on $\theta$ while the others do not.

From Theorem \ref{thm:1} (iv), it follows that the subalgebra
generated by the partial isometries $\{[D\vfi_\theta,D\omega]_t:
t\in \bbbr\}$ is {\bf minimal sufficient}, that is, it is
sufficient and contained in any  sufficient subalgebra. Moreover,
it is invariant under $\sig_t^\omega$. We will denote this
subalgebra by $\iM_\iS$. By Theorem \ref{thm:2}, we have
the decompositions:
\begin{equation}\label{eq:Sdecomp}
D_\theta=D_{\iS,\theta}D_Rz_\iS,\qquad
D_\omega=D_{\iS,\omega}D_Rz_\iS
\end{equation}
where $ D_{\iS,\theta}$,
$D_{\iS,\omega}$ are the densities of the restrictions $\vfi_\theta|_{\iM_\iS}$ 
and $\omega|_{\iM_\iS}$ with respect to the canonical  trace $\tau_\iS$, it will be called the $\iS${\bf -decomposition}.
 The next Theorem shows that each
decomposition of the form (\ref{eq:factortheta}) is given by an invariant
sufficient subalgebra and (\ref{eq:Sdecomp}) is the maximal one.

\begin{thm}\label{T:prop} 
Let us suppose that there is a decomposition $D_\theta=L_\theta R$, with some  
positive operators
$L_\theta$, $R$ in  $\iM$, such that $\supp R=I$ and $R$ commutes with all 
$L_\theta$. Let $\iM_L$ be the von Neumann
algebra generated by $\{ L_\theta:\theta\in\Theta\}$. Then $\iM_L$ is 
sufficient  and invariant under $\sigma_t^\omega$. Moreover,
$$
L_\theta=D_{\iS,\theta}R_0\, ,
$$
where $D_{\iS,\theta}$ is given by (\ref{eq:Sdecomp}) and 
$R_0\in\iM_L$ is a positive element commuting with all $D_{\iS,\theta}$.
\end{thm}

{\it Proof.} We have $D_\omega=\sum_n\lambda_n
D_{\theta_n}=\sum_n\lambda_nL_{\theta_n}R$, hence
$\sum_n\lambda_nL_{\theta_n}$ converges strongly to some positive
operator $L_\omega\in \iM_L$, such that $D_\omega=L_\omega R$. For
$a\in \iM_L$, we get
$$
D_\omega^{it}aD_\omega^{-it}=L_\omega^{it}aL^{-it}_\omega\in \iM_L
$$
and $\iM_L$ is invariant under $\sig_t^\omega$. It follows also
that there is a  density operator $D_{\omega_L}\in\iM_L$ of the restriction $\omega_L:=\omega|_{\iM_L}$, such that
$D_{\omega_L}c=L_\omega$ for some $c\in \iM_L'\cap\iM_L$.
Moreover, it is easy to see that $\iM_\iS\subset\iM_L$, so that
$\iM_L$ is sufficient and the densities of $\vfi_\theta|_{\iM_L}$ satisfy
$$
D_{\theta,L}^{it}c^{it}=[D\vfi_\theta|_{\iM_L},D\omega_L]_tD_{\omega_L}^{it}c^{it}=
[D\vfi_\theta,D\omega]_tL_{\omega}^{it}=L_\theta^{it}
$$
By Theorem \ref{thm:2}, there is a decomposition $D_{\theta,L}=D_{\iS,\theta}D_{R,L}z_L$,
such that $D_{R,L}z_L\in \iM_\iS'\cap \iM_L$. Putting all together, we get
$$
L_\theta=D_{\theta,L}c=D_{\iS,\theta}R_0
$$
where $R_0=D_{R,L}z_Lc\in \iM_\iS'\cap\iM_L$.
\qed

It is easy to see that  the $\iS$-decomposition is, up to a central element 
in $\iM_\iS$, the unique decomposition having the property described in the 
previous theorem.

Keeping the assumptions of Theorem \ref{thm:2}, let us suppose that $\iM$ acts 
on some Hilbert space $\iH$.
The relative commutant $\iM_\iS^c:=\iM_\iS'\cap\iM$ is a countable direct sum 
of factors of type I, hence there is an orthogonal
family of minimal central projections $p_n$ such that $\sum_np_n=1$. 
Therefore, $z_\iS=\sum_n z_np_n$, with some $z_n>0$.
Moreover, there is a decomposition 
\begin{equation}\label{eq:Hdecomp}
\iH=\bigoplus_n\iH^L_n\otimes\iH_n^R,\qquad p_n:\ \iH\to\iH^L_n\otimes \iH^R_n
\end{equation}
such that, up to isomorphism,
\begin{eqnarray*}
\iM^c_\iS &=&\bigoplus_n \bbbc I_{\iH^L_n}\otimes B(\iH^L_n)\\
(\iM^c_\iS)'&=&\bigoplus_nB(\iH^L_n)\otimes \bbbc I_{\iH^R_n}
\end{eqnarray*}
From $D_R\in \iM_\iS^c$ and $D_{\iS,\theta}\in \iM_\iS\subseteq (\iM_\iS^c)'$, 
we have
$$
p_nD_R=c^R_n (1_{\iH^L_n}\otimes D^R_n),\qquad
p_nD_{\iS,\theta}=c^L_n(\theta) (D_n(\theta)\otimes 1_{\iH^R_n}),
$$
where $D^R_n$ is a density operator in $B(\iH^R_n)$, $D_n(\theta)$ is a 
density operator in $B(\iH^L_n)$ and $c_n^R,c_n^L(\theta)>0$.
From  this and (\ref{eq:Sdecomp}), we get the following form of 
the $\iS$-decomposition
\begin{equation}\label{eq:Sdecomp2}
D_\theta=D_{\iS,\theta}D_Rz_\iS=\sum_n z_n p_nD_{\iS,\theta} 
p_nD_R=\sum_n s_n(\theta)D_n(\theta)\otimes D^R_n\, ,
\end{equation}
where $s_n(\theta) \ge 0$ for all $\theta$, $n$. Clearly, this  decomposition 
is unique, up to isomorphisms. It is also clear that $s_n(\theta)=\tau( 
D_\theta p_n)=\vfi_\theta(p_n)$.

In particular, each statistical experiment $(B(\iH),\iS)$, dominated by a
faithful state with finite entropy, defines a decomposition of the form (\ref{eq:Hdecomp})
of the Hilbert space $\iH$, which is up to isomorphisms unique. Note also that 
if the dimension of $\iH$ is finite, then it can be shown from Theorem
\ref{T:prop} that (\ref{eq:Sdecomp2}) gives the  maximal decomposition,  obtained by Koashi 
and Imoto in \cite{koashi}.

\begin{thm}\label{thm:4}
Let $\iK$ and $\iH$ be Hilbert spaces and let $(B(\iH),\iS)$ be a statistical 
experiment,  dominated by a faithful state  $\omega$ with $S(\omega)<\infty$.
Let  $\alpha:\ B(\iK)\to B(\iH)$ be a coarse-graining and let $(B(\iK),\iS_0)$ be the experiment
induced by $\alpha$. Then the following are equivalent.
\begin{enumerate}
\item[(i)] $\alpha$ is sufficient for $(B(\iH),\iS)$.
\item[(ii)] Let (\ref{eq:Hdecomp}) be the decomposition of $\iH$ given by $(B(\iH),\iS)$.
There is a decomposition $\iK=\bigoplus_n\iK^L_n\otimes \iK^R_n$
such that if $q_n:\ \iK\to 
\iK^L_n\otimes\iK^R_n$ is the orthogonal projection, then $\alpha(q_n)=p_n$.
Moreover, 
there are  unitaries $U_n:\ \iK^L_n\to \iH^L_n$ and coarse-grainings 
$\alpha_{n,2}:\ B(\iK^R_n)\to B(\iH^R_n)$ such that
the restriction $\alpha_n:=\alpha|{q_n\iB(\iK) q_n}$ has the form
$$
\alpha_n= \alpha_{1,n}\otimes\alpha_{2,n},\qquad  \alpha_{1,n}(a)=U_naU_n^*,\ 
a\in B(\iK^L_n)
$$

\item[(iii)] Let $D_\theta=D_{\iS,\theta}D_Rz_R$  be the $\iS$-decomposition.
The density $D_{\theta,0}$ of $\vfi_\theta\circ\alpha$ has the form
$$
D_{\theta,0}=L_{\theta,0}\alpha^*(D_Rz_R).
$$
where $L_{\theta,0}\in B(\iK)$ 
is a positive operator satisfying $\alpha(L_{\theta,0})=D_{\iS,\theta}$.
\end{enumerate}
If any of the above conditions is satisfied, then the $\iS_0$-decomposition of the  
densities $D_{\theta,0}$ is
\begin{equation}\label{eq:Sdecomp0}
D_{\theta,0}= \alpha^*_\omega(D_{\iS,\theta})\alpha^*(D_Rz_R)=
\sum_n\vfi_\theta(p_n) U_n^*D_n(\theta)U_n\otimes \alpha_{2,n}^*(D^R_n)
\end{equation}
\end{thm}

{\it Proof.} Let $\iM_\iS\subset B(\iH)$ be generated by 
$\{[D\vfi_\theta,D\omega]_t:\ t\in \mathbb{R}\}$, note that in this case
$\iM_\iS=(\iM_\iS^c)'$ and $\iM_\iS^c=\iM_\iS'$.

Let us denote by $\iN_{\iS_0}\subset B(\iK)$ the subalgebra  generated by
$\{[D(\vfi_\theta\circ\alpha),D(\omega\circ\alpha)]_t,\ t\in \mathbb{R}\}$.
If $\alpha$ is sufficient, then by Theorem \ref{thm:3}, $\iN_{\iS_0}$ is  in the multiplicative 
domain of $\alpha$ and the  restriction $\alpha|{\iN_\iS}$ is a 
*-isomorphism $\iN_{\iS_0}$ onto $\iM_\iS$. Hence, $\iN_{\iS_0}$ has the 
same structure as $\iM_\iS$. Namely, there is an orthogonal family of 
minimal central projections $\{q_n\}$ in $\iN_{\iS_0}$ such that $q_n\iK
=\iK^L_n\otimes \iK^R_n$,
$$
\iN_{\iS_0}=\bigoplus_n B(\iK^L_n)\otimes \bbbc I_{\iK^R_n}, \qquad 
\iN_{\iS_0}'=\bigoplus_n \bbbc I_{\iK^L_n}\otimes B(\iK^R_n)
$$
and $\alpha(q_n)=p_n$.
Moreover, there are unitaries $U_n:\iK_n^L\to \iH^L_n$, such that
if $a\in \iN_{\iS_0}$, $a=\sum_n a_n\otimes I_{\iK^R_n}$ for some $a_n\in B(\iK^L_n)$, then
$\alpha(a)=\sum_n U_na_nU_n^*\otimes I_{\iH^R_n}$.

Let $b\in \iN_{\iS_0}'$, then for $a\in \iM_\iS$,
$$
\alpha(b)a=\alpha(b)\alpha(\alpha^{-1}(a))=\alpha(b\alpha^{-1}(a))=
\alpha(\alpha^{-1}(a)b)=a\alpha(b)
$$
so that $\alpha(b)\in \iM_\iS'$. Consequently, $\alpha(bq_n)=\alpha(b)p_n
\in \iM_\iS'p_n$ and if $b_n\in B(\iK^R_n)$,
then $\alpha_n(I_{\iK^L_n}\otimes b_n)=I_{\iH^L_n}\otimes b'_n$ for 
some $b'_n\in B(\iH_n^R)$. It is clear that the map
$\alpha_{2,n}:\ b_n\mapsto b'_n$ is a coarse-graining $B(\iK^R_n)\to 
B(\iH^R_n)$. We also have
$$
\alpha_n(a_n\otimes b_n)=\alpha_n((a_n\otimes I_{\iK^R_n})(I_{\iK^L_n}
\otimes b_n))=\alpha_n(a_n\otimes I_{\iK^R_n})
\alpha_n(I_{\iK^L_n}\otimes b_n)=U_na_nU_n^*\otimes \alpha_{2,n}(b_n),
$$
hence $\alpha_n=\alpha_{1,n}\otimes \alpha_{2,n}$ and (ii) is proved.

Conversely, let (ii) be satisfied and let  $a\in \iM_\iS$. Then $a=\sum_na_n
\otimes 1_{\iH^R_n}$ and $a=\alpha(b)$ with $b=\sum_nb_n\otimes I_{\iK^R_n}$, 
$b_n=U_n^*a_nU_n$. Clearly, $\alpha(b^*b)=\alpha(b)^*\alpha(b)$, $\alpha(bb^*)
=\alpha(b)\alpha(b)^*$ and therefore $b$ is in the multiplicative domain.
By Theorem \ref{thm:3} (i), $\alpha$ is sufficient for $\iS$ and (i) is proved.

To prove (i) $\to$ (iii), suppose that $\alpha$ is sufficient, then by the first 
part of the proof of (ii), $\iN_{\iS_0}$ is a countable direct sum of type I factors and, 
moreover,  if $\tau_{\iS_0}$ is the canonical trace on $\iN_{\iS_0}$,
then $\tau_{\iS_0}=\tau_\iS\circ\alpha$. We have the $\iS_0$-decomposition
$$
D_{\theta,0}=D_{\iS_0,\theta}D_{R,0}z_{R,0}
$$
where $D_{\iS_0,\theta}$ is the density  of $\vfi_{\theta}\circ\alpha|{\iN_{\iS_0}}$
with respect to $\tau_{\iS_0}$. For $a\in \iN_{\iS_0}$, $\alpha(a)\in\iM_\iS$ and
$$
\tau_{\iS_0}(D_{\iS_0,\theta}a)=\vfi_\theta(\alpha(a))=
\tau_\iS(D_{\iS,\theta}\alpha(a))=
\tau_\iS(\alpha(\alpha^*_\omega(D_{\iS,\theta}))\alpha(a))=
\tau_{\iS_0}(\alpha^*_\omega(D_{\iS,\theta})a),
$$
hence $D_{\iS_0,\theta}=\alpha^*_\omega(D_{\iS,\theta})$ and $\alpha(D_{\iS_0,\theta})
=D_{\iS,\theta}$. Further, let $a\in B(\iK)$, then
$$
\Tr D_{\theta,0}a=\Tr D_\theta\alpha(a)=\Tr \alpha(D_{\iS_0,\theta})D_Rz_R\alpha(a)=
\Tr D_{\iS_0,\theta}\alpha^*(D_Rz_R)a
$$
and (iii) follows, with $L_{\theta,0}=D_{\iS_0,\theta}$.

Conversely, suppose (iii) and let $a\in B(\iK)$, then
$$
\Tr D_{\theta,0}a=\Tr L_{\theta,0}\alpha^*(D_Rz_R)a=\Tr \alpha(a L_{\theta,0})D_Rz_R
$$
On the other hand
$$
\Tr D_{\theta,0}a=\Tr D_\theta\alpha(a)=\Tr \alpha(a)\alpha(L_{\theta,0})D_Rz_R
$$
In particular, by putting  $a=L_{\theta,0}$, we get
$$
\Tr (\alpha(L_{\theta,0}^2)-\alpha(L_{\theta,0})^2)D_Rz_R=0
$$
From this and Schwarz inequality, we get
$$
W^{1/2}(\alpha(L_{\theta,0}^2)-\alpha(L_{\theta,0})^2)W^{1/2}=0
$$
where $W=D_Rz_R$ is positive, $\supp W=1$. Consequently, $\alpha(L_{\theta,0}^2)
=\alpha(L_{\theta,0})^2$, hence
$L_{\theta,0}$ is in the multiplicative domain. Since $D_{\iS,\theta}=
\alpha(L_{\theta,0})$ generates $\iM_\iS$,
this implies that $\alpha$ is sufficient.

It remains only to prove the second half of (\ref{eq:Sdecomp0}), which follows 
easily from (ii). \qed

\begin{coro}\label{coro:cp} 
Let  $\iH$ and $\iK$ be finite dimensional Hilbert
spaces. Let $(B(\iH),\iS)$ be a statistical experiment dominated by a
faithful state $\omega$ and let (\ref{eq:Hdecomp}) be the corresponding
decomposition of $\iH$.  Suppose that
$\alpha:\ B(\iK)\to B(\iH)$ is a   completely
positive map, with the Kraus representation
$\alpha(a)=\sum_i V_iaV_i^*$. Then $\alpha$ is sufficient
for $(B(\iH),\iS)$ if and only if there is a decomposition 
$\iK=\bigoplus_n\iK_n^L\otimes\iK_n^R$ and  
$$
V_i=\sum_n U_n\otimes L_{i,n}
$$
where
$U_n:\ \iK^L_n\to\iH^L_n$ are unitary and $L_{i,n}:\ \iK^R_n\to\iH^R_n$
are linear maps such that $\sum_i L_{i,n}L_{i,n}^*=1_{H_n^R}$.
\end{coro}

{\it Proof.} Note first that $S(\omega)<\infty$, so that the conditions of
Theorem \ref{thm:4} are satisfied.

It is clear that if $V_i$ have the above form, then the
restrictions 
$$\alpha_n=\alpha|B(\iK_n^L\otimes \iK_n^R)=\alpha_{1,n}\otimes\alpha_{2,n},$$
with $\alpha_{1,n}(a)=U_naU_n^*$ and $\alpha_{2,n}(a)=\sum_i L_{i,n}aL_{i,n}^*$. By
Theorem \ref{thm:4} (ii), $\alpha$ is sufficient.

Conversely, if  $\alpha$ is sufficient, then there is a decomposition  
$\iK=\bigoplus_n\iK_n^L\otimes\iK_n^R$ and the corresponding projections 
$q_n$ satisfy $\alpha(q_naq_m)=p_n\alpha(a)p_m$. Consequently
$$
\alpha(a)=\sum_{n,m}p_n\alpha\Big(\sum_{k,l}q_kaq_l\Big)p_m=
\sum_{n,m}p_n\alpha(q_naq_m)p_m=
\sum_i\Big(\sum_np_nV_iq_n\Big)a\Big(\sum_mq_mV_i^*p_m\Big)
$$
Let $V_{i,n}:=p_nV_iq_n$, then $V_{i,n}:\ B(\iK^L_n)\otimes B(\iK^R_n)\to B(\iH^L_n)\otimes B(\iH^R_n)$
and
$$
\sum_i V_{i,n}aV_{i,n}^*=\alpha_n(a), \qquad a\in B(\iK^L_n)\otimes B(\iK^R_n).
$$
By Theorem \ref{thm:4}, there are unitaries $U_n:\ \iK^L_n\to\iH^L_n$ and
coarse-grainings
$\alpha_{2,n}:B(\iK^R_n)\to B(\iH^R_n)$ such that $\alpha_n=\alpha_{1,n}\otimes\alpha_{2,n}$, in fact, 
it is easy to see that $\alpha_{2,n}$ have to be completely positive. This implies that there are linear maps
$K_{i,n}:\ \iK^R_n\to \iH^R_n$, $\sum_iK_{i,n}K_{i,n}^*=1_{\iH_n^R}$, such that
$$
\alpha_n(a)=\sum_i (U_n\otimes K_{i,n})a(U_n\otimes K_{i,n})^*
$$
is another Kraus representation of $\alpha_n$. Hence there are
$\{\mu_{i,j}^n\}$, $\sum_i\mu_{i,j}^n\bar{\mu}_{i,k}^n=\delta_{j,k}$, such that
$V_{i,n}=U_n\otimes\sum_j\mu^n_{i,j}K_{j,n}$. Similarly, there are $\nu_{i,j}$,
$\sum_i\nu_{i,j}\bar\nu_{i,k}=\delta_{j,k}$, such that
$$
V_i=\sum_j\nu_{i,j}(\sum_nV_{j,n})=\sum_nU_n\otimes L_{i,n}
$$
where $L_{i,n}=\sum_{j,k}\nu_{i,j}\mu_{j,k}^nK_{k,n}$.
\qed

As another corollary, we obtain a result previously proved in \cite{koashi}.

\begin{coro} Under the assumptions of Corollary \ref{coro:cp}, suppose that
$\iK=\iH$. Let $D_\theta=\sum_n\vfi_\theta(p_n)D_n(\theta)\otimes D^R_n$ be the $\iS$-decomposition. 
Then $\vfi_\theta\circ\alpha=\vfi_\theta$ for all $\vfi_\theta\in\iS$ if and only if
$$
V_i=\sum_n1_{\iH^L_n}\otimes L_{i,n}
$$
where $\sum_iL_{i,n}L_{i,n}^*=1_{\iH_n^R}$ and $L_{i,n}$ commutes with $D^R_n$ for all $i$, $n$.
\end{coro}

{\it Proof.} Let $\alpha$ satisfy $\vfi_\theta\circ\alpha=\vfi_\theta$ for all $\theta$, then $\alpha$ is obviously
sufficient and by Corollary \ref{coro:cp}, $V_i=\sum_nU_n\otimes L_{i,n}$. On the other hand, by (\ref{eq:Sdecomp0}),
$$
D_\theta=D_{\theta,0}=\sum_n\vfi_\theta(p_n)U_n^*D_n(\theta)U_n\otimes \alpha_{2,n}^*(D^R_n)
$$
and therefore $U_nD_n(\theta)U_n^*=D_n(\theta)$  and
$\alpha_{2,n}^*(D^R_n)=\sum_iL_{i,n}^*D^R_nL_{i,n}=D^R_n$ for all $\theta$ and $n$.
By construction of the $\iS$-decomposition
(\ref{eq:Sdecomp2}), the operators $D_n(\theta)$ generate $B(\iH^L_n)$, hence $U_n=1_{\iH^L_n}$. Moreover, the
operator $D^R_n$ is in the fixed point space of $\alpha_{2,n}^*$ if and only if it commutes with the Kraus operators
$L_{i,n}$ for all $i$, \cite{HJPW}.

The converse statement is obvious.
\qed

\section{Exponential families}

A set of measures $\Pe =\{P_{\theta},\ \theta\in \Theta\}
\abs \mu$ is an {\bf exponential family} if
there are functions $\xi_1,\dots,\xi_m:\ \Theta \to \mathbb{R}$ and
measurable functions
$T_1,\dots,T_m:\ X\to \mathbb{R}$ such that for all $\theta \in \Theta$
$$
\frac{dP_\theta}{d\mu}(x) = \frac{1}{Z(\theta)}
\exp\left(\sum_{i=1}^m \xi_i(\theta)T_i(x)\right)h(x)\,.
$$
In this case, all elements in $\Pe$ are mutually equivalent.

It is immediate from  the factorization criterion  that the statistic
$T=(T_1,\dots,T_m)$ is  sufficient for $\Pe$. Moreover, it is minimal
sufficient if the functions $\{1_{\Pe},\xi_1,\dots,\xi_m\}$ are linearly
independent.

Let $\Pe$ be a family of measures such that the elements are
mutually equivalent. Then $\Pe$ is an  exponential family if and only if the
linear space spanned by the functions
$\{\log \frac{dP}{d\mu},\ P\in \Pe\}$, is finite dimensional.

In the non-commutative case, let us assume that $\omega $ is a state
of the finite dimensional algebra $\iM$ and assume that the density of
$\omega$ is written in the form $\exp H$, $H=H^*\in \iM$. Determine
the states $\vfi_\theta$ by their density
\begin{equation} \label{E:qef}
D_\theta:=\frac{\exp \left(H+\sum_i \xi_i(\theta)a_i\right)}{Z(\theta)},
\end{equation}
where $\xi_1,\dots,\xi_m:\ \Theta \to \bbbr$ are functions, $a_1,
a_2,\dots, a_m$ are self-adjoint operators from $\iM$ and $Z(\theta)$
is for normalization. We call (\ref{E:qef}) {\bf quantum exponential family}
around $\omega$. One can always assume that $\omega(a_i)=0$ in (\ref{E:qef}).

The next example tells us how the exponential family arises.

\begin{ex}
Let $a_1, a_2,\dots, a_m$ be self-adjoint operators from an algebra $\iM$ and
assume that the density of a state $\omega$ is written in the form $\exp H$,
$H=H^*\in \iM$, moreover $\omega(a_i)=0$. If $\Theta$ is a small neighborhood
of $0 \in \bbbr^n$, then minimization of $S(\psi, \omega)$ under the
constraints $\psi(a_i)=\theta_i$ ($\theta=(\theta_1,\theta_2,\dots,\theta_n) \in
\Theta, 1\leq i \leq n$) gives a state $D_\theta$ which is of the form
(\ref{E:qef}) and we arrive at an exponential family.
The functions $\xi_i(\theta)$ are determined by the constraints
$$
\frac{1}{Z(\theta)}\Tr \exp \left(H+\sum_i \xi_i(\theta)a_i\right)a_j=\theta_j,
$$
which has a unique solution if $\theta_j$ are small enough. \qed
\end{ex}

Let $\iM$ be a von Neumann algebra and $\omega$ be a normal state. For $a \in
\iM^{sa}$ define the state $[\omega^a]$ as the minimizer of
\begin{equation}\label{E:min}
\psi \mapsto S(\psi, \omega)-\psi(a).
\end{equation}
If the density of $\omega$ is $e^H$, then the density of  $[\omega^a]$
is nothing else but
$$
\frac{\exp \left(H+ a\right)}{\Tr\exp \left(H+ a\right) },
$$
therefore we can extend the above concept of exponential family as
\begin{equation} \label{E:qqef}
\theta \mapsto \varphi_\theta:=[\omega^{\sum_i \theta_i a_i}],
\end{equation}
where $a_1,a_2,\dots,a_n$ are self-adjoint operators from $\iM$. Note that
the support of the above states is $\supp \omega$. For more details about
perturbation of states, see Chap. 12 of \cite{OP} but here we recall
the analogue of (\ref{E:qef}) in the general case. We assume that the von
Neumann algebra is in a standard form and the representative of $\omega$ is
$\Omega$ from the positive cone. Let $\Delta_\omega$ be the modular operator
of $\omega$ then $\varphi_\theta$ of (\ref{E:qqef}) is the vector state induced
by the unit vector
\begin{equation}
\Phi_\theta:= \frac{\exp \fel \Big(\log \Delta_\omega+ \sum_i \theta_i
a_i \Big) \Omega}{\Big\|\exp \fel\Big(\log \Delta_\omega+ \sum_i \theta_i
a_i\Big) \Omega \Big\|}\,.
\end{equation}
(This formula holds in the strict sense if $\omega$ is faithful, since
$\Delta_\omega$ is invertible in this case. For non-faithful $\omega$
the formula is modified by the support projection.)

In the next theorem $\sigma^\omega_t$ denotes the modular automorphism
group of $\omega$, $\sigma^\omega_t(a)=\Delta_\omega^{\im t}a\Delta_
\omega^{-\im t}$.

\begin{thm} {\bf \cite{petz1986}}\label{T:4}
Let $\iM$ be a von Neumann algebra with a faithful normal state $\omega$
and $\iM_0$ be a subalgebra. For $a_1,a_2,\dots,a_n \in \iA^{sa}$ the
following conditions are equivalent.
\begin{enumerate}
\item[(i)] $\iM_0$ is sufficient for the exponential family (\ref{E:qqef})
\item[(ii)] $\sigma^\omega_t(a_i)\in \iM_0$ for all $t \in \bbbr$ and $1 \leq i \leq n$.
\item[(iii)] For the generalized conditional expectation $E_\omega:\iM\to
\iM_0$ $E_\omega(a_i)=a_i$ holds, $1 \leq i \leq n$.
\end{enumerate}
\end{thm}

Let us denote by $c(\omega,a)$ the minimum in (\ref{E:min}), that is,
$c(\omega,a)=S([\omega^a],\omega)-[\omega^a](a)$. Then
$$
c(\omega,a)=-\log \omega^a(1),
$$
where $\omega^a$ is the positive functional induced by the vector
$\exp \fel \Big(\log \Delta_\omega+ a \Big) \Omega$.
The function $\theta\mapsto c(\omega,\sum\theta_ia_i)$ is analytic and
$$
-\frac{\partial}{\partial\theta_j}c(\omega,
\textstyle{\sum_i}\theta_ia_i)=\vfi_\theta(a_j),
\qquad \hbox{for all}\theta \hbox{\ and\ } j\,.
$$

\begin{thm}\label{thm:exsuff} Let $\iN$, $\iM$ be  von Neumann algebras and let $\alpha:\iN\to
\iM$ be a coarse-graining. Let $\omega$ be a faithful normal state on
$\iM$ and suppose that
$\omega_0:=\omega\circ\alpha$ is also  faithful.
Let $\vfi_\theta$, $\theta\in \Theta$ be the exponential
family $\vfi_\theta=[\omega^{\sum_i\theta_i b_i}]$ for
 $b_1,\dots b_k\in \iM^{sa}$.
Then $\alpha$ is sufficient for $(\iM,\vfi_\theta)$ if and only if
$b_i=\alpha(a_i)$, $i=1,\dots,n$ for some $a_i\in \iN^{sa}$ and
\begin{equation}\label{E:exsuff}
\vfi_\theta\circ\alpha=\left[\omega_0^{\sum_i\theta_ia_i}\right].
\end{equation}
\end{thm}

{\it Proof.}  Let $\alpha$ be sufficient for $(\iM,\vfi_\theta)$ and let
$$
\iN_1=\{ a\in \iN, \alpha^*_{\omega}\circ\alpha(a)=a\}=\{a\in \iN_{\alpha},
\alpha(\sig_t^{\omega_0}(a))=\sig_t^\omega(\alpha(a))\}.$$
Then $\alpha(\iN_1)$
is a sufficient subalgebra and therefore
$\sigma_t^{\omega}(b_j)\in
\alpha(\iN_1)$ for all $t$, $j=1,\dots,k$, in particular, $b_i=\alpha(a_i)$,
$a_i\in \iN_1$.  Let $a(\theta)=\sum_j\theta_ja_j$
and consider the expansion
\begin{eqnarray*}
[D\vfi_\theta,D\omega]_t&=&[D\omega^{\alpha(a(\theta))},D\omega]_t\\
&=&\sum_{n=0}^\infty i^n\int_0^tdt_1\dots
\int_0^{t_{n-1}}dt_n\sig_{t_n}^\omega(\alpha(a(\theta)))...\sig_{t_1}^\omega
(\alpha(a(\theta)))\\
&=& \sum_{n=0}^\infty i^n\int_0^tdt_1\dots
\int_0^{t_{n-1}}dt_n\alpha(\sig_{t_n}^{\omega_0}(a(\theta)))...\alpha(
\sig_{t_1}^{\omega_0}(a(\theta)))\\
&=& \alpha([D\omega_0^{a(\theta)},D\omega_0]_t)
\end{eqnarray*}

On the other hand, $\alpha$ is sufficient, therefore
$[D\vfi_\theta,\omega]_t\in
\alpha(\iN_\alpha)$ and
$$
\alpha([D\vfi_\theta\circ\alpha,D\omega_0]_t)=[D\vfi_\theta,D\omega]_t
$$
As $\alpha$ is invertible on $\iN_\alpha$, it follows that
$[D\vfi_\theta\circ\alpha,D\omega_0]_t=[D[\omega_0^{a(\theta)}],D\omega_0]_t$
and (\ref{E:exsuff}) follows.

Conversely, let $b_i=\alpha(a_i)$ for some $a_i\in \iN$ and
suppose (\ref{E:exsuff}), then
$$
\frac{\partial}{\partial \theta_j}c(\omega_0,a(\theta))=
-[\omega_0^{a(\theta)}](a_j)=-\vfi_\theta(\alpha(a_j))=
\frac{\partial}{\partial\theta_j}c(\omega,\alpha(a(\theta)))
$$
for all $\theta$ and $j$. Putting $\theta=0$, it follows that
$c(\omega_0,a(\theta))=c(\omega,\alpha(a(\theta)))$ for all $\theta$.
Hence
$$
S(\vfi_\theta,\omega)=c(\omega,\alpha(a(\theta)))+
\vfi_\theta(\alpha(a(\theta)))=c(\omega_0,a(\theta))+\vfi_\theta\circ\alpha
(a(\theta))=S(\vfi_\theta\circ\alpha,\omega\circ\alpha)
$$
and $\alpha$ is sufficient.
\qed

{\it Remark.}
Note that in case $\iM=B(\iH)$, $\dim \iH=n$, the condition (\ref{E:exsuff})
reads
$$
\alpha(\log \alpha^*(D_{\theta})-\log \alpha^*(D_{\omega_0}))=
\log D_\theta-\log D_\omega,
$$
where $\alpha^*$ is the dual of $\alpha$ with respect to $\<A,B\>=\Tr
A^*B$.
This condition is  known to be equivalent to sufficiency of $\alpha$.
\qed

\begin{coro}
Let $\iM$ be a von Neumann algebra with a faithful normal state $\omega$,
$\iM_0$ a commutative subalgebra and  (\ref{E:qqef}) the
exponential family for $a_1,a_2,\dots,a_n \in \iM^{sa}$. Then
 $\iM_0$ is sufficient for the exponential family if and only if
$a_1,\dots,a_n\in \iM_0$ and
$$
\vfi_\theta(a)=\omega\big(\exp\big(\ssum_i\theta_ia_i\big)a\big)\quad a\in\iM
$$
\end{coro}

{\it Proof.}  Let $\iM_1$ be the subalgebra generated by $\sig_t^\omega(a_i)$,
$t\in R$, $i=1,\dots,n$. Then $\iM_1$ is sufficient, by Theorem
\ref{T:4}.
Let $E: \iM\to \iM_1$ be the $\omega$ preserving conditional expectation, then
$E$ preserves all $\vfi_\theta$, by sufficiency (Theorem \ref{thm:1} (iv)).
If $\iM_0$ is sufficient, then $\iM_1\subseteq \iM_0$, hence  $\iM_1$ is
commutative. Let $\omega_0$, $\vfi_{\theta,0}$ be the restriction of $\omega$,
$\vfi_\theta$ to $\iM_1$, then by Theorem \ref{thm:exsuff},
$$
\vfi_{\theta,0}=[\omega_0^{\sum\theta_ia_i}]=
\omega_0(\exp(\ssum\theta_i a_i)\,\cdot\, )
$$
It follows that for $a\in \iM$,
$$
\vfi_\theta(a)=\vfi_{\theta,0}(E(a))=\omega_0(\exp(\ssum\theta_ia_i)E(a))=
\omega(\exp(\ssum\theta_ia_i)a)
$$

Conversely, let $a_1,\dots,a_n\in \iM_0$ and let
$\vfi_\theta=\omega(\exp\sum\theta_ia_i)\,\cdot\, )$, then the restriction
of $\vfi_\theta$ to $\iM_0$ is the exponential family
$[\omega_0^{\sum\theta_ia_i}]$ and $\iM_0$ is sufficient, by Theorem
\ref{thm:exsuff}.
\qed

\section{Strong subadditivity of entropy}

Let $\iH=\iH_A\otimes\iH_B\otimes\iH_C$ and let $\omega_{ABC}$ be a
normal state on $B(\iH)$ with restrictions $\omega_B, \omega_{AB}$
and $\omega_{BC}$. The von Neumann entropies satisfies the {\bf
strong subadditivity}
\begin{equation}\label{eq:ssa}
S(\omega_{ABC})+S(\omega_B)\le S(\omega_{AB})+S(\omega_{BC})\,,
\end{equation}
which was obtained by Lieb and Ruskai \cite{LR}. A concise proof
using the Jensen operator inequality is contained in \cite{petz1986b}
and \cite{N-P} is a didactical presentation of the same ideas.
As we want to investigate the case of equality mostly, we  suppose
below that all the involved entropies are finite. The case of equality
was studied in several papers recently but always restricted to finite 
dimensional Hilbert spaces \cite{HJPW, MP}. Our aim now is to allow 
infinite dimensional spaces.

The strong subadditivity is equivalent to
\begin{equation}\label{eq:lim}
S(\omega_{AB},\omega_A\otimes\omega_B) \le S(\omega_{ABC},\omega_A
\otimes\omega_{BC}),
\end{equation}
which is a consequence of monotonicity of the relative entropy.
Clearly, the equality in (\ref{eq:ssa}) is equivalent to equality in
(\ref{eq:lim}) which means that $B(\iH_A)\otimes B(\iH_B)$ is a 
sufficient subalgebra for the states $\omega_{ABC}$ and $\omega_A
\otimes\omega_{BC}$. Our results on factorization apply.

\begin{thm} Let $\omega_{ABC}$ be a faithful normal state on $B(\iH)$ such that
the von Neumann entropy $S(\omega_{ABC})$ is finite and
$$
S(\omega_{ABC})+S(\omega_B)=S(\omega_{AB})+S(\omega_{BC}).
$$
Then there is a decomposition $\iH_B=\bigoplus_n \iH_{nB}^L\otimes \iH_{nB}^R$
such that
\begin{equation}\label{eq:ssadecomp} 
\omega_{ABC}=\sum_n\omega_B(p_n)D^L_n\otimes D^R_n\, ,
\end{equation}
where $D^L_n\in B(\iH_A)\otimes B(\iH_{nB}^L)$ and $D^R_n\in B(\iH^R_{nB})
\otimes B(\iH_C)$ are density operators
and $p_n\in B(\iH_B)$ are the orthogonal projections $\iH_B\to \iH_{nB}^L
\otimes\iH_{nB}^R$.
\end{thm}

{\it Proof.} Equality in the strong subadditivity is equivalent to 
sufficiency of the subalgebra $B(\iH_A\otimes\iH_B)\otimes 
\mathbb{C}1_C$ for $(B(\iH),\iS)$ where $\iS:=\{\omega_{ABC},\omega_A
\otimes\omega_{BC}\}$, and the latter is  equivalent to 
$$
[D\omega_{ABC},D\omega_A\otimes\omega_{BC}]_t=[D\omega_{AB},D\omega_A
\otimes\omega_B]_t\otimes1_C
$$
for all $t$.  Let $\iN_B\subset B(\iH_B)$ be the  subalgebra
$$
\iN_B=\{ b\in B(\iH_B):\ \sigma_t^{\omega_{BC}}(b\otimes I_C)=
\sigma_t^{\omega_B}(b)\otimes I_C \mbox{\ for every\ }t\in \bbbr\}\,.
$$
Then in follows from the above equality and the cocycle condition that
$$
[D\omega_{AB},D(\omega_A\otimes\omega_B)]_t\in B(\iH_A)\otimes \iN_B\quad 
\mbox{for all\ } t
$$
and therefore $B(\iH_A)\otimes\iN_B\otimes \mathbb{C}1_C$ is sufficient for
$\iS$. Since $\omega_A\otimes\omega_{BC}$ is faithful,
$S(\omega_A\otimes\omega_{BC})<\infty$ and clearly dominates $\iS$,
moreover, the subalgebra is invariant under $\sigma_t^{\omega_A\otimes
\omega_{BC}}$, we
have by Theorem \ref{thm:2} that there is a decomposition
$$
D_{ABC}=(D_L\otimes 1_C)(1_A\otimes D_R),
$$
where $D_L\in B(\iH_A)\otimes\iN_B,\ D_R\in\iN_B'\otimes B(\iH_C)$ are density
operators.

On the other hand, $\iN_B$ is invariant under $\sigma^{\omega_{B}}_t$,
therefore $S(\omega_{B}|_{\iN_B})\le S(\omega_{B})< \infty$. Similarly as in
Section \ref{sec:factorization}, we obtain a decomposition  $\iH_B=\bigoplus_n 
\iH_{nB}^L\otimes \iH_{nB}^R$
such that
$$
\iN_B=\bigoplus_n B(\iH_{nB}^L)\otimes \mathbb{C}1_{\iH^R_{nB}}\qquad \iN'_B=
\bigoplus \mathbb{C}1_{\iH_{nB}^L}\otimes B(\iH_{nB}^R)
$$
and (\ref{eq:ssadecomp}) follows.\qed

The structure (\ref{eq:ssadecomp}) of the density matrix $\omega_{ABC}$ 
is similar to the finite dimensional situation discussed in 
\cite{HJPW, MP}, however the direct sum decomposition may be infinite.

The theorem is stated under the condition of faithfulness of $\omega_{ABC}$. 
It would be worthwhile to weaken this condition. When $\omega_{ABC}$ is 
pure the strong subadditivity reduces to
$$
S(\omega_{AC})\le S(\omega_{A})+S(\omega_{C}),
$$
which is simply the subadditivity. The equality holds here if $\omega_{AC}=
\omega_{A}\otimes \omega_{C}$. Since the purification of a product state is
a product vector, we have the product structure (\ref{eq:ssadecomp}) 
(without the summation over $n$). 
Note that this kind of states were discussed in \cite{MN}. 

The decomposition (\ref{eq:ssadecomp}) has a continuous version formulated
in terms of direct integrals (see \cite{Pdir} for references about the 
direct integral of fields of Hilbert spaces and operators or \cite{Schw}). 
Let $(X,\mu)$ be a measure space. Assume that for $x \in X$ density matrices
$D^L (x)\in B(\iH_A)\otimes B(\iH^L(x))$ and $D^R (x) \in B(\iH^R(x))
\otimes B(\iH_C)$ such that $\iH^L(x)$ and $\iH^R(x)$ are measurable
fields of Hilbert spaces and the operator fields $D^L (x)$ and $D^R (x)$
are measurable as well, $x \in X$. Given a probability density $p(x)$
on $X$
\begin{equation}\label{E:integral} 
\omega_{ABC}:=\dint p(x)D^L (x)\otimes D^R (x)\,d\mu(x)
\end{equation}
is a density on the Hilbert space $\iH_A \otimes \iH_B \otimes \iH_C$,
where
$$
\iH_B:=\dint \iH^L(x)\otimes \iH^R(x)\, d\mu(x)\, .
$$
Then $B(\iH_A)\otimes B(\iH_B)$ is a sufficient subalgebra for 
the states $\omega_{ABC}$ and $\omega_A \otimes\omega_{BC}$. If the
measure $\mu$ is not atomic, then $S(\omega_{ABC})=\infty$.

\section{Appendix}

\subsection*{Dual mapping}
Let $\iM_1$ and $\iM_2$ be von Neumann algebras and let $\sigma:\iM_1
\to \iM_2$ be a coarse-graining. Suppose that a normal state $\vfi_2$
is given and $\vfi_1:=\vfi_2\circ \sigma$ is normal as well.

We assume that both von Neumann algebras are in a standard form and
the representative of $\vfi_i$ is $\Phi_i$ from the positive cone.
From the modular theory we know that
$$
p_i:= \overline{J_i\iM_i\Phi_i}
$$
is the support projection of $\vfi_i$ (i=1,2).

The dual $\aa: p_2\iM_2 p_2 \to  p_1\iM_1 p_1$ of $\sigma$ is
is characterized by the property
\begin{equation}
\langle A_1, J_1\aa(A_2)\rangle =
\langle \sigma(A_1), J_2 A_2 \rangle
\end{equation}
(see Prop. 8.3 in \cite{OP}). The dual of the embedding of a subalgebra into
an algebra is called generalized conditional expectation \cite{AC}.

\end{document}